# Can the GPC standard eliminate consent banners in the EU?

Sebastian Zimmeck, Wesleyan University szimmeck@wesleyan.edu
Harshvardhan J. Pandit, AI Accountability Lab, Trinity College Dublin me@harshp.com
Frederik Zuiderveen Borgesius, Radboud University frederikzb@cs.ru.nl
Cristiana Teixeira Santos, Utrecht University c.teixeirasantos@uu.nl
Konrad Kollnig, Maastricht University konrad.kollnig@maastrichtuniversity.nl
Robin Berjon, IPFS Foundation & Supramundane Agency robin@berjon.com

## Abstract

In the EU, the General Data Protection Regulation and the ePrivacy Directive mandate consent for the use of personal data for the purpose of behavioural advertising and tracking technologies. However, the ubiquity of consent banners has led to widespread consent fatigue and questions about the effectiveness of these mechanisms in protecting data subjects' data. To simplify digital laws and make the EU more competitive, the EU Commission recently proposed the Digital Omnibus, introducing a new Article 88b GDPR to express data subjects' choices in a technical way. While the Digital Omnibus is under legislative negotiation, California residents and residents of other US states can already exercise their rights via Global Privacy Control (GPC), a privacy signal to automatically broadcast a legally binding opt-out request to websites. In light of the Digital Omnibus, we evaluate to which extent GPC can be adapted to the EU legal framework to reduce consent banners, mitigate consent fatigue, and improve data protection for EU users.

GPC is based on a technical specification, currently being standardised at the World Wide Web Consortium. By sending a GPC signal, data subjects can express their refusal or withdrawal of consent under the GDPR to the use of their personal data for cross-context ad targeting and, in some cases, to express their objection under the GDPR against the use of their data for such purposes. Our evaluation identifies friction between the GPC specification and current EU data protection law. In the longer term, it would be possible for the EU legislator to amend EU laws, as proposed in the current Digital Omnibus, in such a way that internet users can use automated signals to express choices about personal data use and online tracking. In the shorter term, websites and companies who conduct online tracking can already honour GPC. Websites can choose to show no or fewer consent banners when they see a user's GPC signal. Also in the shorter term, regulators can clarify how websites should interpret automated signals like GPC in various situations.

## Keywords

# 1 Introduction

On 19 November 2025, the European Commission presented its Digital Omnibus simplification package.[1] Motivated by international competitive pressures and worries about limited economic growth in the EU,[2] the Digital Omnibus aims to increase the EU's international competitiveness by reducing the burden arguably imposed on business by various laws governing the online space in the European Union, including the 2016 General Data Protection Regulation (GDPR)[3] and the 2009 ePrivacy Directive (ePD).[4] A key innovation of the Digital Omnibus is the proposed Article 88b GDPR that requires technical standards for machine-readable indications of data subjects' choices. It is in this context that we explore how Global Privacy Control (GPC),[5] being standardised at the World Wide Web Consortium (W3C), can be used in EU data protection law. GPC enables users to automatically broadcast an opt-out request to websites. We focus on GPC as it has existing adoption of significance, is legally enforced in California and other states in the US, and is supported by multiple stakeholder groups. In particular, we ask whether GPC can help to overcome EU citizens' *consent fatigue* by eliminating, or, at least, limiting consent banners.

Data subjects' consent to personal data processing has a long history. In the EU, *informed consent*—or *notice and choice,* as it is known in the US—plays a large role in the EU's fundamental rights to privacy and data protection.[6] However, many see informed consent as ineffective for protecting privacy.[7] The GDPR applies broadly to the processing of personal data and imposes a variety of obligations on *data controllers*, i.e., natural or legal persons who determine the data processing purposes and its means.[8] For some data processing purposes, such as behavioural advertising, which typically uses cookies for tracking and profiling of users across websites, the ePD and GDPR require data controllers to obtain a data subject's informed consent for lawful processing.[9] While the GDPR sets the requirements for the validity of informed consent, the ePD, in its Article 5(3), requires any party that stores or accesses information on a user's device to obtain that user's informed consent, which functionally governs the use of cookies and other tracking technologies.

---

[1] European Commission, 'Digital Omnibus Regulation Proposal' (2025) https://digital-strategy.ec.europa.eu/en/library/digital-omnibus-regulation-proposal accessed 21 April 2026.

[2] M. Draghi, 'The future of European competitiveness: Report by Mario Draghi' (2024) https://commission.europa.eu/topics/competitiveness/draghi-report_en accessed 21 April 2026.

[3] Regulation (EU) 2016/679 of the European Parliament and of the Council of 27 April 2016 on the protection of natural persons with regard to the processing of personal data and on the free movement of such data, and repealing Directive 95/46/EC (General Data Protection Regulation) (2016), OJ 2016 L 119/1.

[4] Directive 2002/58/EC of the European Parliament and of the Council of 12 July 2002 concerning the processing of personal data and the protection of privacy in the electronic communications sector (Directive on privacy and electronic communications) (2002), OJ 2002 L201/37, as amended by Directive 2006/24/EC (Data Retention Directive), and Directive 2009/136/EC (Citizen's Rights Directive).

[5] Global Privacy Control, https://globalprivacycontrol.org/ accessed 21 April 2026.

[6] Article 7 and 8 Charter of Fundamental Rights of the European Union (2012), OJ C326/391.

[7] C. Utz et al., '(Un)informed Consent: Studying GDPR Consent Notices in the Field' (2019) Proc. ACM SIGSAC Conf. Computer and Communications Security (CCS); M. Nouwens et al., 'Dark Patterns after the GDPR: Vanishing, Reconfigured, or Persistent?' (2022) Proc. Privacy Enhancing Technologies (PoPETs); F.J. Zuiderveen Borgesius, 'Improving privacy protection in the area of behavioural targeting', Information Law Series, Kluwer Law International 2015, https://hdl.handle.net/11245/1.434236 accessed 21 April 2026.

[8] Article 4(7) GDPR.

The requirements of the GDPR and ePD are usually implemented on websites in the form of banners that act as consent interfaces while also serving to meet other obligations regarding transparency and exercise of data rights.[10] The design, usability, and effect of these interfaces have been studied extensively.[11] Studies have shown that existing consent interfaces often do not comply with EU law.[12] Consent interfaces often implement deceptive practices—*deceptive patterns,* also called *dark patterns*—that nudge or trick users into consenting.[13] But even when sites do not implement dark patterns, users are tired of repetitive decision-making on nearly every site they visit and experience consent fatigue.[14]

In contrast to this burdensome implementation of the EU's legal requirements which fosters consent fatigue due to per-site interaction, residents in California and a number of other states in the US are increasingly benefitting from a single *global setting* in their browser that websites must legally support, such as by using the GPC setting in their browser.[15] After web users express their choice in their browsers,[16] those automatically broadcast a legally binding signal, asserting their right to opt out of the sale or sharing of personal information. The legal enforceability of GPC distinguishes it from prior efforts such as the Do Not Track (DNT) signal from 2009,[17] which was also being standardised at the W3C but was not legally enforceable anywhere.[18] The offices of the California, Colorado, and Connecticut attorneys general have already started enforcing this legal obligation.[19] The growing use and enforcement of GPC combined with its standardisation at the W3C are thus likely to make GPC an integral part of the web. A recent amendment to the California Consumer Privacy Act (CCPA), the California Opt Me Out Act per California Assembly Bill 566,[20] requires a business that develops or maintains a browser to include functionality to send an opt-out preference signal by 1 January 2027.

---

[9] Article 29 Data Protection Working Party, 'Opinion 06/2014 on the Notion of Legitimate Interests of the Data Controller under Article 7 of Directive 95/46/EC' (2014), pp. 46-47, https://ec.europa.eu/justice/article-29/documentation/opinion-recommendation/files/2014/wp217_en.pdf accessed 21 April 2026.

[10] In particular, Articles 13 and 14 GDPR regarding information to be made available to the data subject, and Articles 18 and 21 GDPR regarding the right to restrict processing via an exercised objection.

[11] For a comprehensive overview see E. Birrell et al., 'SoK: Technical Implementation and Human Impact of Internet Privacy Regulations' (2024) IEEE Symposium on Security and Privacy (SP), pp. 673-696.

[12] C. Santos, N. Bielova and C. Matte, 'Are cookie banners indeed compliant with the law?' (2020) Technology and Regulation, pp. 91-135, https://doi.org/10.71265/g317tv72 accessed 21 April 2026.

[13] See, e.g., C.M. Gray et al., 'An Ontology of Dark Patterns Knowledge: Foundations, Definitions, and a Pathway for Shared Knowledge-Building' (2024) ACM Proc. CHI Conference on Human Factors in Computing Systems (CHI), Article 289, pp. 1-22, https://doi.org/10.1145/3613904.3642436 accessed 21 April 2026.

[14] B. van der Sloot, 'Editorial' (2024) European Data Protection Law Review, Volume 10, Issue 1, pp. 1-8, https://edpl.lexxion.eu/article/edpl/2024/1/3/display/html accessed 21 April 2026; F.J. Zuiderveen Borgesius, 'Behavioural Sciences and the Regulation of Privacy on the Internet' in A-L Sibony and A. Alemanno (eds.), Nudge and the Law: A European Perspective, pp. 179-207, Hart Publishing (2015), https://arxiv.org/abs/2511.20637 accessed 21 April 2026.

[15] Global Privacy Control, https://globalprivacycontrol.org/ accessed 21 April 2026.

[16] S. Zimmeck et al., 'Generalizable Active Privacy Choice: Designing a Graphical User Interface for Global Privacy Control' (2024) Proc. Privacy Enhancing Technologies (PoPETs), https://doi.org/10.56553/popets-2024-0015 accessed 21 April 2026.

[17] California Online Privacy Protection Act (CalOPPA), Cal. Bus. & Prof. Code §22575-22579.

[18] Compliance with GPC is required in California since January 2021 (https://x.com/AGBecerra/status/1354850758236102656), Colorado since July 2024

As GPC functions as an automated signal to communicate the user's wishes, in some situations, EU websites receiving a GPC signal do not have to show interfaces, such as banners, or ask users to make a decision. This aspect of GPC conveying the user's specific wish or decision without requiring a banner interface has the potential to mitigate the effects of consent fatigue and reduce deceptive patterns. As GPC is technically a simple signalling mechanism, it can be implemented on every platform that uses the web as the underlying communication protocol. Thus, for example, GPC can be used for apps on smartphones or IoT devices and can be configured to all sites or apps, to some, or to none.[21]

In this paper, we explore the question to which extent GPC can be applied in the context of data subjects automating decisions regarding their privacy and data protection rights as provided by laws in the EU. In doing so, we identify the potential but also limitations of GPC within the current EU framework. We make recommendations to improve the development and specification of GPC to assist in implementing it fully within the EU. In particular, we take into account the Commission's proposed Article 88b GDPR, which aims to make technical signals enforceable. To assist with the development of Article 88b GDPR and broader regulatory frameworks, we provide recommendations to lawmakers based on the ongoing standardisation and the existing legal enforceability of GPC in various states in the US.

---

(https://coag.gov/uoom/), Connecticut since January 2025 (https://portal.ct.gov/ag/sections/privacy/the-connecticut-data-privacy-act), New Jersey since July 2025 (https://www.njconsumeraffairs.gov/ocp/Pages/NJ-Data-Privacy-Law-FAQ.aspx), and Oregon since January 2026 (https://www.doj.state.or.us/consumer-protection/id-theft-data-breaches/privacy/). All accessed 21 April 2026.

[19] Office of the California Attorney General, 'Attorney General Bonta Announces Settlement with Sephora as Part of Ongoing Enforcement of California Consumer Privacy Act' (2022) https://oag.ca.gov/news/press-releases/attorney-general-bonta-announces-settlement-sephora-part-ongoing-enforcement, Office of the California Attorney General, 'Attorney General Bonta Announces Largest CCPA Settlement to Date, Secures $1.55 Million from Healthline.com (2025) https://oag.ca.gov/news/press-releases/attorney-general-bonta-announces-largest-ccpa-settlement-date-secures-155, Office of the Connecticut Attorney General, 'Connecticut, California and Colorado Announce Joint Investigative Privacy Sweep' (2025) https://portal.ct.gov/ag/press-releases/2025-press-releases/connecticut-california-and-colorado-announce-joint-investigative-privacy-sweep. All accessed 21 April 2026. For an evaluation of websites' compliance with GPC see Hausladen et al., 'Websites' Global Privacy Control Compliance at Scale and over Time' (2025) USENIX Security, pp. 5837-5856, https://www.usenix.org/system/files/usenixsecurity25-hausladen.pdf accessed 21 April 2026.

[20] California State Legislature, 'AB-566 California Consumer Privacy Act of 2018: opt-out preference signal' (2025) https://leginfo.legislature.ca.gov/faces/billNavClient.xhtml?bill_id=202520260AB566 accessed 21 April 2026.

[21] Throughout this paper we refer to *websites* or just *sites* with the understanding that what we discuss also applies to mobile apps and other web-based software.

Our arguments are structured as follows: section 2 provides a brief history of technical privacy controls on the web, section 3 explains what GPC is, section 4 evaluates the application of GPC under EU law, in section 5 we discuss our findings and give suggestions for further research, and section 6 contains brief conclusions.

# 2 A Brief History of Privacy Controls on the Web

Over the years, there have been various attempts to move beyond the current manual informed consent implementations. Looking back at the history of these attempts, particularly, their successes and failures, is instructive for evaluating the potential of GPC adoption in the EU as well as the adoption of automated and machine-readable choice architectures more broadly.

The first major effort to automate users' privacy choices was the Platform for Privacy Preferences Project (P3P), which was standardised at the W3C in the early 2000s.[22] P3P was an ambitious effort to allow websites to declare their data collection practices in a machine-readable format. Users could, for example, configure their browsers to automatically accept or reject cookies based on whether a site's practices matched their predefined privacy settings. However, P3P faced challenges; its XML-based policy format was complex for websites to implement, and browser support was limited and inconsistent. It failed to gain widespread adoption.[23] There were also usability issues with P3P user agents being perceived as annoying by some users.[24] Ultimately, P3P was abandoned.

After P3P, Do Not Track (DNT) emerged as a simpler signal.[25] Standardised at the W3C from 2009 onwards, DNT allowed users to indicate their preference to permit or prohibit being tracked when accessing websites. While most major browsers implemented the DNT setting, its broad effect was negligible. In particular, legal backing for the standard was limited. A 2013 amendment to the California Online Privacy Protection Act (CalOPPA) required websites to *disclose* how they respond to DNT signals.[26] Notably, the amendment does not require them to respect the signal. Further, the standard left the definition of "tracking" up to individual companies and regulators, which led the advertising industry to largely ignore the signal, claiming a lack of consensus and legal mandate.[27]

[22] L.F. Cranor, 'Web Privacy with P3P', O'Reilly & Associates 2002.
[23] J.R. Reidenberg et al., 'Disagreeable Privacy Policies: Mismatches between Meaning and Users' Understanding' (2015) 30 Berkeley Technology Law Journal 39, https://ir.lawnet.fordham.edu/faculty_scholarship/619 accessed 21 April 2026.
[24] L.F. Cranor, P. Guduru and M. Arjula, 'User Interfaces for Privacy Agents' (2004) https://lorrie.cranor.org/pubs/privacy-bird-20040802.pdf accessed 21 April 2026.
[25] J.R. Mayer and J.C. Mitchell, 'Third-Party Web Tracking: Policy and Technology' (2012) IEEE Symposium on Security and Privacy (SP), pp. 413-427.
[26] International Association of Privacy Professionals, 'What Do the New Disclosure Requirements Under CalOPPA Mean for Your Business?' (2013) https://iapp.org/news/a/what-do-the-new-disclosure-requirements-under-caloppa-mean-for-your-busines accessed 21 April 2026.
[27] Future of Privacy Forum, 'Thank you for Visiting AllaboutDNT.com' (2013) https://fpf.org/thank-you-for-visiting-allaboutdnt-com/ accessed 21 April 2026.

The major point of contention was DNT's default setting. The advertising industry particularly criticised Microsoft's decision to enable DNT by default in its Internet Explorer browser as not being an indication of the user's preferences but rather a paternalistic decision by Microsoft as browser vendor.[28] The only case for DNT to have legal enforceability is a recent isolated German state court decision against LinkedIn, though, only because LinkedIn had claimed to respect DNT in its privacy policy.[29] Overall, at this point it is no longer worthwhile to support DNT or base enforcement actions on it as browser vendors have been removing DNT from browser settings and implementations. Of note, for its Firefox browser Mozilla specifically points to GPC as a successor in its DNT deprecation page.[30]

In parallel to user-facing signals the advertising industry has been developing its own technical standards to manage and propagate consent signals throughout the online advertising ecosystem. Most notably, the Interactive Advertising Bureau (IAB) has been developing its Transparency and Consent Framework (TCF) to represent user choices and transmit them to ad networks and other third parties using Real-Time Bidding (RTB) for tracking-based advertising. The TCF has faced significant legal challenges, in particular, as the Court of Justice of the European Union (CJEU) upheld the Belgian Data Protection Authority's findings that it violates the GDPR.[31] In response the IAB is updating the TCF and, independently, is integrating it into its Global Privacy Platform (GPP), which aims to consolidate the TCF with similar IAB frameworks in other regions of the world. TCF and GPP are intended for propagating users' privacy choices across the advertising ecosystem and rely on consent banners or automated signals, such as GPC, as user-facing mechanisms. As such, they do not help to mitigate consent fatigue. Additionally, they are often implemented in ways that result in harms to individual's rights and contravene the principles of the GDPR.[32]

Various frameworks for fine-grained privacy choices have been proposed. Prominent among those are Advanced Data Protection Control (ADPC),[33] which is intended to be aligned with the EU's GDPR, and the Data Rights Protocol (DRP),[34] which is designed for comprehensively communicating data rights, not just the opt-out right. Consenter,[35] which is certified under the German Telecommunications and Telemedia Data Protection Act (TTDSG, 2021),[36] is a privacy manager that allows users to set preferences through a browser extension, which is then communicated to websites in a manner that allows automated acceptance and rejection of consent. The IEEE 7012-2025 Standard for Machine Readable Personal Privacy Terms allows individuals to express their privacy requirements, such as whether to share data with third parties as contractual terms between websites and users.[37]

---

[28] B. Lynch, 'Do Not Track in the Windows 8 Setup Experience' (2012) https://blogs.microsoft.com/on-the-issues/2012/08/07/do-not-track-in-the-windows-8-setup-experience/ accessed 21 April 2026.

[29] Landgericht Berlin, 1.12.2022, 52 O 157/22.

[30] Mozilla, 'How do I turn on the Do Not Track feature?' (2026) http://mzl.la/WL6fUP accessed 21 April 2026.

[31] CJEU Case C-604/22 (*IAB Europe*, 2024); M. Veale, M. Nouwens and C. Santos, 'Impossible Asks: Can the Transparency and Consent Framework Ever Authorise Real-Time Bidding After the Belgian DPA Decision?' (2022) Technology and Regulation, pp. 12-22, https://doi.org/10.26116/techreg.2022.002 accessed 21 April 2026.

[32] C. Matte, C. Santos and N. Bielova, 'Purposes in IAB Europe's TCF: Which Legal Basis and How Are They Used by Advertisers?' (2020) Annual Privacy Forum, pp. 163-185, https://dl.acm.org/doi/10.1007/978-3-030-55196-4_10 accessed 21 April 2026.

[33] S. Gerben & S. Human, 'Advanced Data Protection Control (ADPC)' (2021) https://www.dataprotectioncontrol.org/ accessed 21 April 2026. For a comparison of GPC with ADPC, see S. Human et al., 'Data Protection and Consenting Communication Mechanisms: Current Open Proposals and Challenges' (2022) IEEE European Symposium on Security and Privacy Workshops (EuroS&PW), pp. 231-239, https://doi.org/10.1109/EuroSPW55150.2022.00029 accessed 21 April 2026.

[34] Consumer Reports, 'Data Rights Protocol' (2026) https://datarightsprotocol.org/ accessed 21 April 2026.

While frameworks like ADPC offer a more fine-grained and comprehensive approach to give users agency over their privacy preferences, they also create concerns regarding browser fingerprinting risks.[38] A user's fine-grained privacy preferences could be used as additional information to uniquely (re-)identify and track users across sites. Overcoming the fingerprinting risk is not insurmountable, but it will take time to evolve user agent implementations, including questions of usability, alongside legal mandates that enforce broad adoption of such frameworks. Article 25 GDPR, which obliges data protection by design and by default, provides a legal mechanism to move towards a fine-grained framework.[39] In addition to the browser fingerprinting risk, a comprehensive framework could also have a larger attack surface for corporate lobbying.[40] The European Data Protection Board (EDPB) has acknowledged the potential of automated technical signals, noting that DNT's adoption was hampered by lack of legal enforceability.[41] The recent proposal of the Digital Omnibus provides the first explicit use of automated technical signals to manage consent and privacy choices within the EU.[42] Under the proposal, the EU wants to develop standards that can assist data subjects in communicating their privacy preferences to reduce the burden of interactions and decisions they need to make on websites. Another relevant proposal in this context is the Digital Fairness Act which aims to address issues regarding deceptive patterns, choice, and consumer harms.[43] However, both proposals are still in the early stages of the lawmaking process, and the associated standards development processes will also take a few years. While we have some considerations for the integration of automated signals into the future legal framework of the EU, we mainly focus on the use of GPC in the EU data protection legal framework as it stands today.

---

[35] Law & Innovation Technology, 'Consenter' (2026) https://www.consenter.eu/ accessed 21 April 2026.
[36] Dienste zur Einwilligungsverwaltung nach §26 TDDDG (2025).
[37] IEEE SA, 'IEEE 7012-2025: IEEE Approved Draft Standard for Machine Readable Personal Privacy Terms' (2026) https://standards.ieee.org/ieee/7012/7192/ accessed 21 April 2026.
[38] Browser or device fingerprinting is a tracking technique in which the user's information and configurations, such as browser version, language used, and screen size, are used to create a unique "fingerprint" for identifying the user across websites. See, e.g., P. Laperdrix et al., 'Browser Fingerprinting: A Survey' (2020) 14(2) ACM Trans. Web, Article 8, pp. 1-33, https://doi.org/10.1145/3386040 accessed 21 April 2026.
[39] M. Veale, R. Binns, and J. Ausloos, 'When data protection by design and data subject rights clash' (2018) 8(2) International Data Privacy Law, pp. 105-123, https://doi.org/10.1093/idpl/ipy002 accessed 21 April 2026.
[40] Corporate Europe Observatory, How corporate lobbying undermined the EU's push to ban surveillance ads (2022) https://corporateeurope.org/en/2022/01/how-corporate-lobbying-undermined-eus-push-ban-surveillance-ads accessed 21 April 2026.
[41] European Data Protection Board, Letter by Anu Talus Chair of the European Data Protection Board (2023) https://www.edpb.europa.eu/system/files/2023-12/edpb_letter_out2023-0099_donottrack_en.pdf accessed 21 April 2026.
[42] European Commission, 'Digital Omnibus Regulation Proposal' (2025) https://digital-strategy.ec.europa.eu/en/library/digital-omnibus-regulation-proposal accessed 21 April 2026.
[43] European Commission, 'Proposal for a Directive/Regulation of the European Parliament and of the Council on strengthening consumer protection in the digital environment (Digital Fairness Act)' (2025),

# 3 What Is GPC?

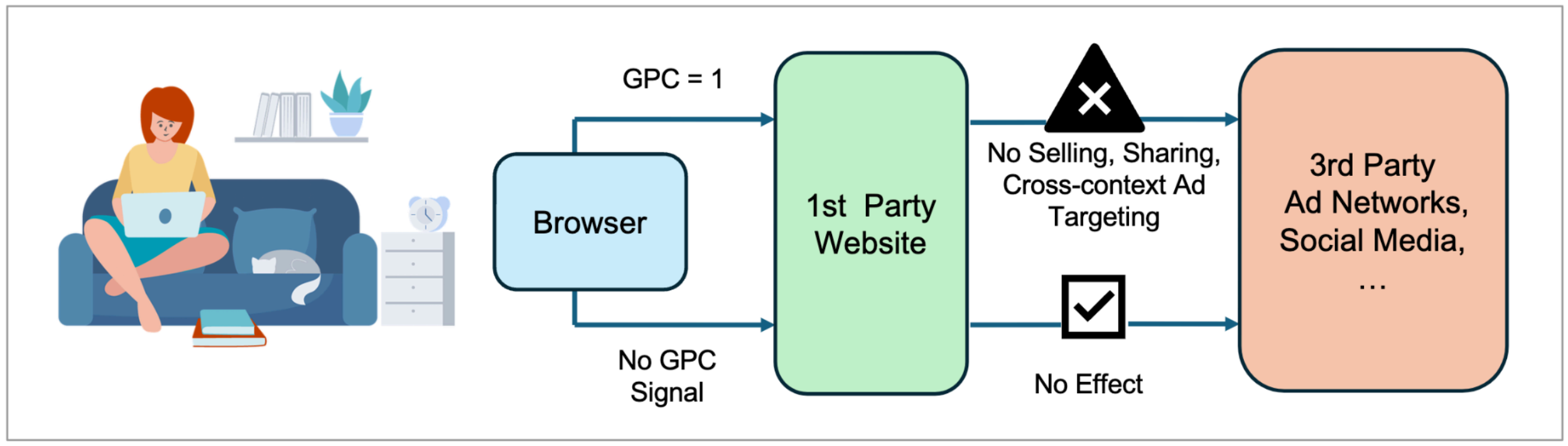

Figure 1. Overview of an active GPC signal sent by the user's browser (GPC = 1), which, under California's CCPA, indicates to the recipient that the consumer does not want their personal data to be sold or shared with third parties or, used for cross-context ad targeting. The GPC signal is attached to every web request to a website. A GPC signal can be configured to be communicated to all sites or just some[44].

## 3.1 Legal Support

GPC is a technical signal that expresses individuals' privacy and data processing preferences to internet services, including apps and websites, that they interact with. These preferences, as defined in the draft specification at the W3C, are that "their data not be sold to or shared with any party other than the one the person intends to interact with, or to have their data used for cross-context ad targeting."[45] The specification's origin is the opt-out principle, on which the CCPA and other US state privacy laws are based and according to which users can be tracked until they exercise their right to opt out (Fig. 1). Thus, the specification provides for GPC signals to convey users' preferences to prevent certain types of processing, but it does not provide a mechanism for users to approve, i.e., consent, to the processing of personal data. Owing to this conceptual alignment, GPC has achieved legal recognition and enforceability in five US states and more are likely to adopt GPC in the future.[46]

---

https://ec.europa.eu/transparency/documents-register/detail?ref=Ares(2025)5829481 accessed 21 April 2026.

[44] The GPC specification alludes to such a mechanism in §2 under do-not-sell-or-share preference and provides an example implementation via the OptMeowt browser extension in §5 of its Legal and Implementation Considerations Guide. See S. Zimmeck, P. Snyder, J. Brookman and A Zucker-Scharff, 'Global Privacy Control (GPC) W3C Editor's Draft 3 April 2026' https://w3c.github.io/gpc/; A. Zucker-Scharff, J. Brookman and S. Zimmeck, 'Global Privacy Control (GPC) Legal and Implementation Considerations Guide' https://w3c.github.io/gpc/explainer. All accessed 21 April 2026.

[45] S. Zimmeck, P. Snyder, J. Brookman and A. Zucker-Scharff, 'Global Privacy Control (GPC) W3C Editor's Draft 3 April 2026' https://w3c.github.io/gpc/ accessed 21 April 2026.

[46] Compliance with GPC is required in California since January 2021 (https://x.com/AGBecerra/status/1354850758236102656), Colorado since July 2024

GPC is an abstract signal that can support different legal interpretations depending on the applicable law. In that sense, GPC is jurisdiction-agnostic. Legislators and regulators have broad leeway to define the meaning of a GPC signal under the laws they enact and enforce, i.e., define what it means to *sell* or *share* or perform *cross-context ad targeting*. The GPC specification originated against the background of emerging US state privacy laws and in the context of the web platform. However, it is principally applicable to any privacy and data protection laws worldwide, including the GDPR,[47] and on different platforms, such as smart cars or TVs. While the GPC specification does not make any claims as to what should happen if no GPC signal is being received, "[t]he specification should not be interpreted as an endorsement of the opt-out model of regulation [...] or a rejection of other models based on consent or data minimization." In other words, while it does not ascribe meaning to the absence of a GPC signal, the GPC specification does not stand in the way of GPC being applied in consent-based jurisdictions. Our analysis is based on the GPC specification as of its current version.[48]

## 3.2 GPC's Main Characteristics

***Preference* and Legal Interpretation.** Per the GPC specification, with a *preference* individuals specify whether they want their data to be processed. Preferences are mapped to specific legal requirements. For example, in California, the use of GPC is interpreted as exercising the opt-out right as defined in the CCPA and its regulations.[49] The GPC specification does not prescribe a fixed interpretation for the meaning of a GPC signal in terms of which legal obligations and restrictions it invokes and instead requires the relevant legislators, regulators, and other authorities responsible for online privacy and data protection to formulate this interpretation. For jurisdictions that adopt GPC their laws and regulations and interpretations of these determine the application of the GPC signal and how the user's preferences must be implemented as part of implementers' obligations. In the absence of regulatory, legal, or other requirements websites can interpret an expressed GPC preference as they find most appropriate for the given person, particularly as considered in light of the person's privacy expectations, context, and cultural circumstances.[50]

---

(https://coag.gov/uoom/), Connecticut since January 2025 (https://portal.ct.gov/ag/sections/privacy/the-connecticut-data-privacy-act), New Jersey since July 2025 (https://www.njconsumeraffairs.gov/ocp/Pages/NJ-Data-Privacy-Law-FAQ.aspx), and Oregon since January 2026 (https://www.doj.state.or.us/consumer-protection/id-theft-data-breaches/privacy/). All accessed 21 April 2026.

[47] R. Berjon, 'GPC under the GDPR' (2021) https://berjon.com/gpc-under-the-gdpr/; H. Pandit, 'GPC + GDPR: will it work?' (2021) https://harshp.com/research/blog/gpc-gdpr-can-it-work. All accessed 21 April 2026.

[48] As the W3C GPC specification is evolving in the standardisation process, our analysis here is based on the version dated 3 April 2026. See S. Zimmeck, P. Snyder, J. Brookman and A. Zucker-Scharff, 'Global Privacy Control (GPC) W3C Editor's Draft 3 April 2026' https://w3c.github.io/gpc/ accessed 21 April 2026.

[49] §1798.120 CCPA, §7025 and §7026 CCPA Regulations.

[50] See S. Zimmeck, P. Snyder, J. Brookman and A. Zucker-Scharff, 'Global Privacy Control (GPC) W3C Editor's Draft 3 April 2026' https://w3c.github.io/gpc/ accessed 21 April 2026.

In jurisdictions where the communication of the user's preferences creates specific obligations and rights, the use of GPC is intended to exercise these.[51] Illustrative examples of such legal effects are provided in the GPC specification and its accompanying Legal and Implementation Considerations Guide,[52] among others, for the CCPA and GDPR. However, the examples in the specification and in the guide are *non-normative*, i.e., not binding and provided as guidance pending authoritative sources such as laws, regulatory guidance, or case law that establish concrete interpretation of GPC in specific jurisdictions. As an example, we can consider GPC's interpretation under the CCPA.

§1798.140(ad)(1) CCPA defines *selling* as communicating a consumer's personal information by a business to a third party for monetary or other valuable consideration.[53] Thus, for example, if a website integrates a free analytics library to understand how its site performs in different regions of the world and in the process discloses consumers' IP addresses to the analytics service, it would engage in selling of personal information. In exchange for free performance statistics the site operator discloses consumers' IP addresses. §1798.140(ah)(1) CCPA defines *sharing* as communicating a consumer's personal information by the business to a third party for cross-context behavioral advertising, whether or not for monetary or other valuable consideration. Further, §1798.140(k) CCPA defines *cross-context behavioral advertising* as the targeting of advertising to a consumer based on the consumer's personal information obtained from the consumer's activity across businesses and distinctly branded websites and apps. Thus, examples of personal data sharing under the CCPA are the integration of targeting cookies or pixels.

**Focus on Single Controller.** Generally, the GPC specification gives users a means to limit their interaction to a single controller, i.e., the entity represented by the website the user intentionally visits and wants to interact with, also called *first party*.[54] This limitation aligns with the experiences of individuals in consumer interactions, for example, when they buy goods in a brick-and-mortar store and interact with the store's staff for buying the goods, returning them, or exercising other rights. Such interactions respect individuals' privacy in the form of contextual integrity.[55] GPC envisions a similar simplified and streamlined experience for users regarding their online privacy as most web users only want to interact with the first party site they visit and not with ad networks or other third parties commonly integrated by websites for ad purposes.[56]

---

[51] Ibid.
[52] A. Zucker-Scharff, J. Brookman and S. Zimmeck, 'Global Privacy Control (GPC) Legal and Implementation Considerations Guide' https://w3c.github.io/gpc/explainer accessed 21 April 2026.
[53] See also S. Zimmeck and K. Alicki, 'Standardizing and Implementing Do Not Sell' (2020) Workshop on Privacy in the Electronic Society (WPES), pp. 15-20, https://dl.acm.org/doi/10.1145/3411497.3420224 accessed 21 April 2026.
[54] The name for GPC during its initial development phase was Signal of a Preference for One Controller (SPOC) to indicate that the user wishes to interact only with the first party whose website they are visiting and not with the third parties integrated on that site that sell or share the user's personal data or perform cross-context ad targeting.
[55] H. Nissenbaum, 'Privacy as Contextual Integrity' (2004) 79(1) Wash. L. Rev., pp. 119-158.
[56] J. Turow et al., 'Americans Reject Tailored Advertising and Three Activities that Enable It' (2009), https://ssrn.com/abstract=1478214; C. Farronato, A. Fradkin and T. Lin, 'Designing Consent: Choice Architecture and Consumer Welfare in Data Sharing' (2025) NBER Working Paper 34025, https://www.nber.org/papers/w34025. All accessed 21 April 2026.

Thus, the boundary between first and third party activities is determined by the *context* of these activities.

According to the W3C Privacy Principles,[57] which the GPC specification references for defining key terms, a context is a physical or digital environment in which people interact with other actors, and which the people understand as distinct from other contexts. An *actor* is an entity that a person can reasonably understand as a single "thing" they are interacting with, which can be people or collective entities like companies, associations, or governmental bodies.[58] The actor that makes or delegates decisions about the content and data processing on this origin or site is known as the web page's first party.[59] A third party is any actor other than the person visiting the website or the first parties they expect to be interacting with.[60] Sharing data between different contexts of a single company can be a privacy violation, just as if the same data were shared between unrelated actors.[61] Notably, the boundary between first and third party activities does not depend on domains or ownership as those do not align with people's privacy expectations and also increase further anti-competitive tendencies and market concentration.[62]

**Technical Differences Between GPC and DNT.** While GPC's legal support (section 3.1) is a major difference to DNT, there are also technical differences. Unlike a DNT signal, which can have either of two values—(1) users permit tracking or (2) users prohibit tracking—GPC has only a single value that indicates a preference to prevent data selling or sharing as well as cross-context ad targeting. If GPC is turned "on" or "activated," the browser communicates this single value with each request to the website. If GPC is turned "off" or "deactivated," no GPC signal is sent to the website. Thus, GPC provides for only 1 bit of new information with which a user could be (re-)identified across websites, which reduces its fingerprinting risks.

While DNT solely relies on headers, the GPC specification allows user agents not only to send GPC signals via GPC headers, but websites can also query the value of the GPC header field via a client-side script (GPC DOM property).[63] Many websites that implement GPC make use of this second option because it allows them to retrieve the GPC status faster compared to receiving it via a sent GPC header. This speedup is relevant for online advertising where websites want to determine as soon as possible whether they are allowed to collect data, track a user, or should display an ad based on the current website context instead.

---

[57] R. Berjon and J. Yasskin, 'Privacy Principles W3C Statement 15 May 2025' https://www.w3.org/TR/privacy-principles/ accessed 21 April 2026. The Privacy Principles are published as a *Statement*, which is the most mature state for documents other than technical specifications to be considered a W3C standard. They are produced by the W3C's Technical Architecture Group (TAG) to ensure architectural coherence of the web and apply to the entire web, including uses of GPC.
[58] Ibid.
[59] Ibid.
[60] Ibid.
[61] Ibid.
[62] R. Berjon, 'GPC W3C GitHub repository, Clarifications and anchoring #108' (2025) https://github.com/w3c/gpc/pull/108#issuecomment-3681219422 accessed 21 April 2026.
[63] S. Zimmeck, P. Snyder, J. Brookman and A Zucker-Scharff, 'Global Privacy Control (GPC) W3C Editor's Draft 3 April 2026' https://w3c.github.io/gpc/ accessed 21 April 2026.

Different from DNT, the GPC specification also provides for an optional GPC Support Resource.[64] With the GPC Support Resource, a website can publicly declare that it respects GPC. It can do so by hosting a /.well-known/gpc.json file relative to its origin server's URL. For example, The New York Times declares that it respects GPC at its site under https://www.nytimes.com/.well-known/gpc.json. The GPC Support Resource is a declaration by a site that it supports GPC but it does not prove that the site actually does so and does not create any additional obligations.

**GPC's Integration into the Online Ad Ecosystem.** Compared to the IAB's TCF, which, from a user's perspective, is "invisible" and generally not directly controllable by users, GPC operates as a user-facing and user-controlled signal that integrates with downstream compliance mechanisms, such as the IAB's TCF (which the IAB is integrating in its GPP) or Google's Restricted Data Processing (RDP).[65] Third parties whose code runs on a first-party website can also detect and act upon the GPC signal directly. However, significant data processing occurs outside the user's browsing session. For instance, personal data collected by a first party during a site visit may be shared server-side with Meta using the Conversions API.[66] In such scenarios, GPC signals are to be "translated" into server-side mechanisms by the first party to control sending the personal data. If that would not happen, downstream recipients would unknowingly process data in violation of the user's rights, creating a compliance gap between the user's browser and the backend data supply chain. GPC is technology-neutral and can apply to (prevent) any activity that involves selling, sharing, or cross-context ad targeting, including but not limited to the use of cookies, browser fingerprinting, and other web technologies.

## 3.3 GPC Adoption

GPC can be manually activated in the settings of web browsers that support GPC (e.g., Firefox) or by installing a browser or browser extension that comes with GPC enabled by default (e.g., Brave or DuckDuckGo).[67] As the CCPA was amended by the California Opt Me Out Act, all browsers for use by California residents will be required to implement GPC by 1 January 2027.[68] Figure 2 shows how users can turn on GPC in two browsers.[69]

---

[64] Ibid.

[65] Google, 'Helping advertisers comply with the U.S. states' privacy laws in Google Ads' https://support.google.com/google-ads/answer/9614122 accessed 21 April 2026.

[66] Meta, 'Conversions API' https://developers.facebook.com/docs/marketing-api/conversions-api/ accessed 21 April 2026.

[67] Mozilla Support, 'Global Privacy Control' (2026) https://support.mozilla.org/en-US/kb/global-privacy-control; DuckDuckGo, 'Global Privacy Control (GPC) in DuckDuckGo' (2026) https://duckduckgo.com/duckduckgo-help-pages/privacy/gpc; P. Snyder, 'Global Privacy Control, a new Privacy Standard Proposal' (2023) https://brave.com/web-standards-at-brave/4-global-privacy-control/. For enabling GPC by default under US state law, see S. Zimmeck, 'Remarks on the Relevance of Privacy Expectations for Default Opt-out Settings' (2026) https://doi.org/10.48550/arXiv.2603.15705. All accessed 21 April 2026.

[68] California State Legislature, 'AB-566 California Consumer Privacy Act of 2018: opt-out preference signal' (2025) https://leginfo.legislature.ca.gov/faces/billNavClient.xhtml?bill_id=202520260AB566 accessed 21 April 2026.

[69] In a usability survey, 46/49 (94%) of the participants expressed interest in a GPC setting and would turn it on if it were available in their browser. See S. Zimmeck et al., 'Usability and Enforceability of Global

A recent study reports the use of GPC resulting in fewer initial third-party tracking cookies being deployed on websites and a further reduction when refusing them.[70] As of 5 April 2026, about 388.000 sites support GPC,[71] among which are Amazon, the National Football League, and Spotify. While there is not yet any study on the effectiveness of GPC in the EU, Figure 3 illustrates how three sites respond to GPC when accessed from Germany.

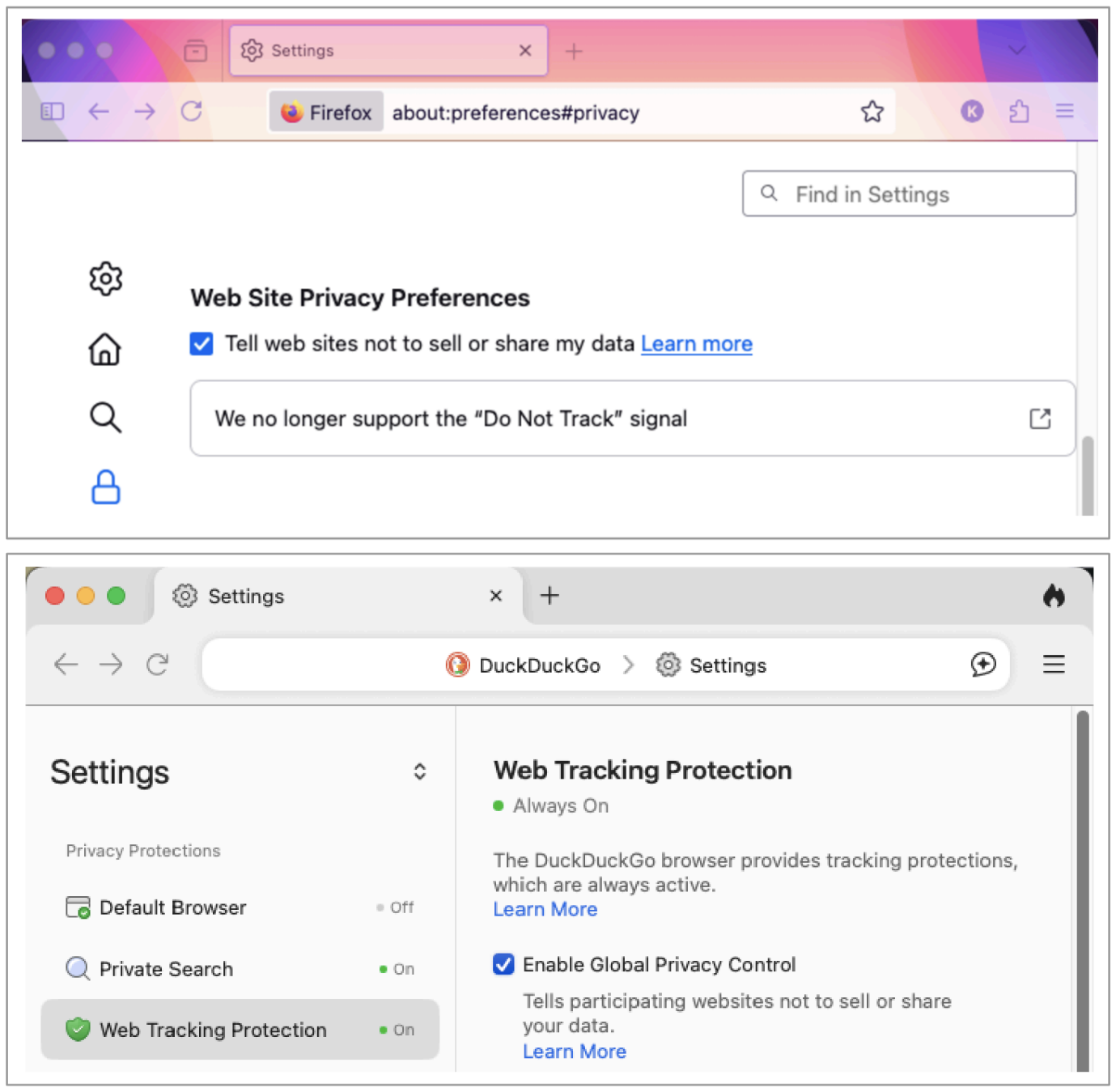

Figure 2. GPC can be enabled from the settings of many privacy-friendly browsers, including Firefox (top) and DuckDuckGo (bottom).

---

Privacy Control' (2023) Proc. Privacy Enhancing Technologies (PoPETs) https://doi.org/10.56553/popets-2023-0052 accessed 21 April 2026.

[70] See A. Rasaii et al., 'Intractable Cookie Crumbs: Unveiling the Nexus of Stateful Banner Interaction and Tracking Cookies' (2025) Proc. Privacy Enhancing Technologies (PoPETs) https://doi.org/10.56553/popets-2025-0138 accessed 21 April 2026: "Our findings reveal that around 50% of websites send at least one intractable cookie, with the majority set to expire after more than 10 days. In addition, enabling the Global Privacy Control (GPC) signal initially reduces the number of intractable cookies by 30% on average, with a further 32% reduction possible on subsequent visits by rejecting the banners."

[71] GPC SUP, Check if a site supports Global Privacy Control, https://gpcsup.com/. A site is counted as supporting GPC if it has a GPC Support Resource (section 3.2).

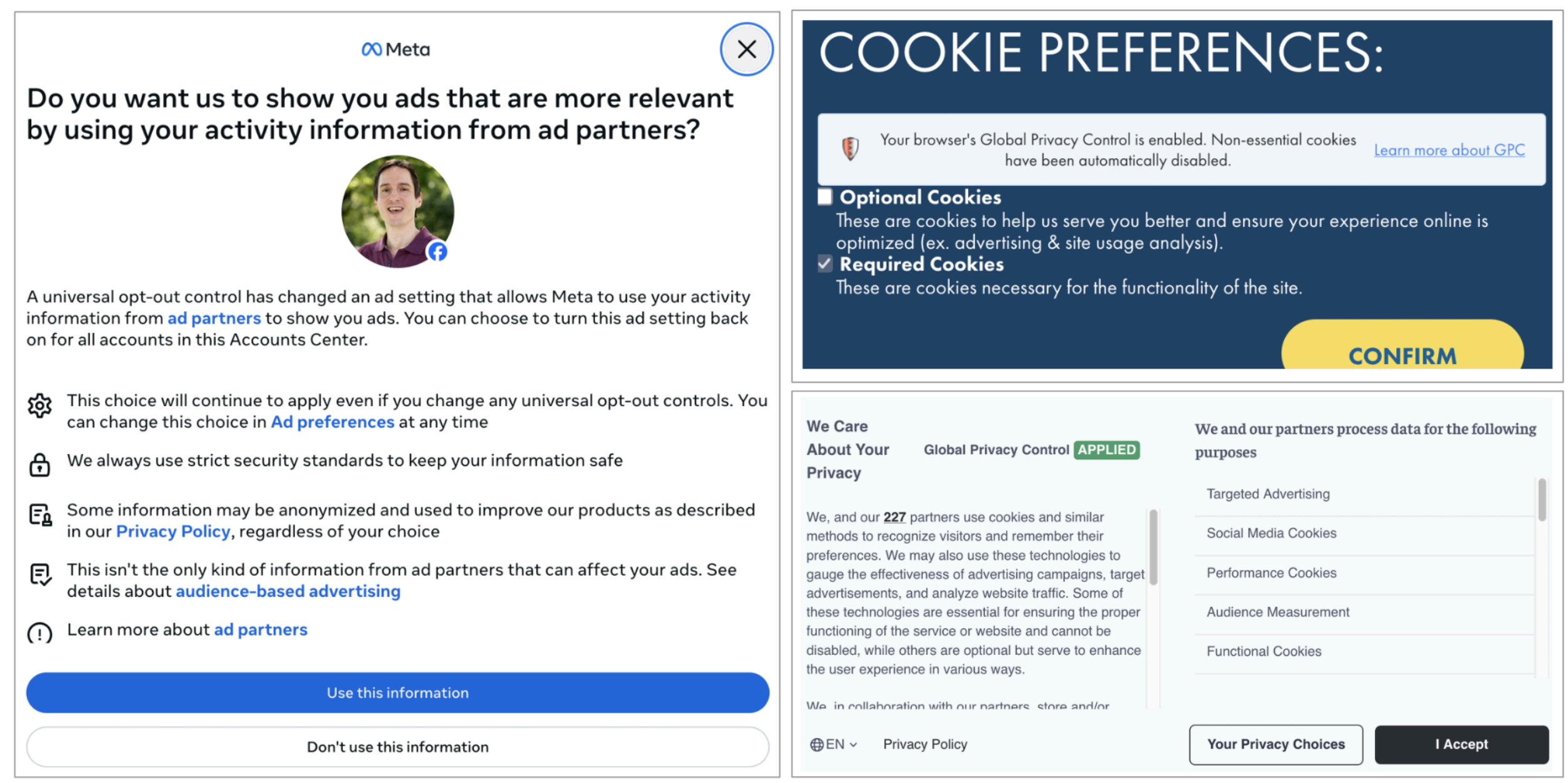

Figure 3. On facebook.com (left) Meta showed a banner with the explanation: "A universal opt-out control has changed an ad setting that allows Meta to use your activity information from ad partners to show you ads. You can choose to turn this ad setting back on for all accounts in the Accounts Center" (December 2024). dutchbros.com (top right) disabled optional cookies (December 2025), and arstechnica.com (bottom right) confirmed: "Global Privacy Control Applied" (December 2025).

While GPC can be implemented on any platform that uses the web as its underlying communication protocol, it is so far only adopted in web browsers and on websites. Support on mobile app platforms or other platforms is not yet available.[72] This lack of adoption is noteworthy as some platforms already implement privacy choice mechanisms, e.g., Apple for iOS via its App Tracking Transparency (ATT) framework.[73] However, these mechanisms are based on contractual agreements between developers and platform companies and not on statutory mandates. The coexistence of statutory and contractual opt-out mechanisms would not be ideal. Instead, there should be one consistent mechanism that ensures the intended privacy effects, is directly controlled by users, and is binding on all actors.[74]

## 3.4 GPC Standardisation at the W3C

GPC is being standardised in the W3C's Privacy Working Group.[75] Currently, the specification is a *Working Draft*. As is the case for all W3C standards it is being drafted through a consensus-based mechanism. The next stages for the GPC specification are *published* and

---

[72] S. Zimmeck et al., 'Exercising the CCPA Opt-out Right on Android: Legally Mandated but Practically Challenging' (2026) Proc. Privacy Enhancing Technologies (PoPETs) https://arxiv.org/abs/2407.14938 accessed 21 April 2026.
[73] K. Kollnig et al., 'Goodbye tracking? Impact of iOS App Tracking Transparency and Privacy Labels' (2022) Proc. ACM Conference on Fairness, Accountability, and Transparency, pp. 508-520, https://dl.acm.org/doi/10.1145/3531146.3533116 accessed 21 April 2026.
[74] S. Zimmeck et al., 'Exercising the CCPA Opt-out Right on Android: Legally Mandated but Practically Challenging' (2026) Proc. Privacy Enhancing Technologies (PoPETs) https://arxiv.org/abs/2407.14938 accessed 21 April 2026.
[75] W3C, Privacy Working Group, https://www.w3.org/groups/wg/privacy/ accessed 21 April 2026.

*ratified*. The W3C's standardisation process deviates from those of other standards organisations in that the W3C requires a standard to have implementation, deployment, and adoption happen *before* ratification in order to ensure that the standard reflects practicality and can be implemented effectively. In fact, it is a pre-condition for a specification to become a W3C standard that it already has two meaningful real-world implementations.[76] GPC meets the criteria for ratification and is being maintained as a draft while remaining editorial work is carried out. Once the GPC specification has been standardised, implementations must adhere to the published standard to be conformant.

The Privacy Principles specify the role of the browser as a *user agent*, which is intended to act as an intermediary representing the user on the web.[77] Notably, the user agent is expected to help users to automate tasks that they wish to carry out. The user agent is also expected to act in users' interests as a trustworthy agent. The Privacy Principles give browsers extensive fiduciary duties as agents of their users.

The Privacy Principles describe how global choices, such as GPC, should be implemented. The description is worth quoting in full:

> When an opt-out mechanism exists, it should preferably work with a global opt-out mechanism. Conceptually, a global opt-out mechanism is an automaton operating as part of the user agent. It is equivalent to a robot that would carry out a person's instructions by pressing an opt-out button (or a similar expression of the person's rights) with every interaction that the person has with a site. (For instance, the person may be objecting to processing based on legitimate interest, withdrawing consent to specific purposes, or requesting that their data not be sold or shared.) The user is effectively delegating the expression of their opt-out to their user agent, which helps rectify automation asymmetry. The Global Privacy Control (GPC) is a good example of a global opt-out mechanism.

As the quote makes clear, GPC is intended to automate an existing legal process and thus requires legal support and enforceability. Thus, sending a GPC signal can be seen as the data subject delegating to the browser the task of systematically and authoritatively expressing their intent of not having their data sold or shared with anyone except the single controller they want to interact with as well as not having their data used for cross-context ad targeting purposes.

# 4 Application of GPC under EU Data Protection Law

In this section, we evaluate the extent to which GPC fits EU data protection law. We focus on the current state of the GDPR and ePD as the key laws regulating the processing of personal data, including selling or sharing, and the use of information from the user's terminal device, respectively. We then explore the applicability of GPC to the proposed Article 88b introduced in the GDPR amendments under the Digital Omnibus.[78]

---

[76] W3C, W3C Process Document, https://www.w3.org/policies/process/ accessed 21 April 2026 ("adequate implementation experience").

[77] R. Berjon and J. Yasskin, 'Privacy Principles W3C Statement 15 May 2025' https://www.w3.org/TR/privacy-principles/ accessed 21 April 2026.

## 4.1 Scope and Extent of GPC

**Personal Data and Data Subject.** The GPC specification relates to *personal information*, which can be interpreted as *personal data* under the GDPR.[79] Similarly, the specification's terminology of *person* and *user* is analogous to the *data subject* under the GDPR and to the *subscriber* or *user* in the ePD.[80]

**Actor's Roles.** Under the GDPR, obligations stem from an actor's role, which include for example, that of a *data controller*,[81] *data processor*,[82] or *third party*.[83] Similarly, the GPC specification refers to entities receiving and implementing GPC signals as *parties*, with a further distinction between *first party* and *third party*. Neither the GPC specification nor its accompanying guide document prescribes how such roles should be interpreted or how they align with the roles defined by the GDPR.[84]

We interpret *first party* to align with the *data controller.*[85] Under the GDPR, the controller is the party that determines the purposes and means of the processing of personal data.[86] The GDPR also allows multiple controllers to jointly determine the purposes and means of processing, in which case they are considered to be *joint controllers*.[87]

The GDPR defines a *third party* as an entity other than the data subject, controller, processor, or person who, under the direct authority of the controller or processor, is authorised to process personal data.[88] The GPC specification defines a third party as "any party other than the one the person intends to interact with."[89] Since both the GDPR and GPC define *third parties* through exclusion, the determination of which entity is a third party depends on an *a priori* understanding of the entity acting as the first party (GPC) or as the controller, processor, etc. (GDPR).

---

[78] European Commission, 'Digital Omnibus Regulation Proposal' (2025) https://digital-strategy.ec.europa.eu/en/library/digital-omnibus-regulation-proposal accessed 21 April 2026.
[79] Article 4(1) GDPR.
[80] Article 4(1) GDPR; Article 5(3) ePD.
[81] Article 4(7) GDPR.
[82] Article 4(8) GDPR.
[83] Article 4(10) GDPR.
[84] S. Zimmeck, P. Snyder, J. Brookman and A. Zucker-Scharff, 'Global Privacy Control (GPC) W3C Editor's Draft 3 April 2026' https://w3c.github.io/gpc/; A. Zucker-Scharff, J. Brookman and S. Zimmeck, 'Global Privacy Control (GPC) Legal and Implementation Considerations Guide' https://w3c.github.io/gpc/explainer. All accessed 21 April 2026.
[85] Article 4(7) GDPR.
[86] Ibid.
[87] Article 26 GDPR. See also CJEU Case C-210/16 (*Wirtschaftsakademie Schleswig-Holstein*, 2018*),* Case C-40/17 (*FashionID*, 2019) and Case C-604/22 (*IAB TCF,* 2024).
[88] Article 4(10) GDPR.
[89] §2 definitions in the GPC specification.

The GDPR further defines a *data processor* as an entity that processes personal data on behalf of a controller.[90] An example could be a cloud provider that stores personal data for the controller. While the GPC specification does not define the role of a processor, it references the W3C Privacy Principles' definition of a *data processor*, which is similar to the notion of a data processor under the GDPR and to which the effects of GPC are not intended to apply.[91] Based on this understanding, we do not interpret a GDPR processor as a third party under the GPC given its role as an authorised entity acting on behalf of a controller. To summarise, we assume the first party under GPC corresponds to a controller under the GDPR, that GDPR processors are similar to GPC processors, and that third parties under the GDPR correspond to third parties under GPC.[92]

**Legal Effects of GPC.** The legal effects of GPC are determined by legislators, regulators, and other authorities under the applicable law of their respective jurisdiction. The GPC specification includes two broad definitions of these legal effects,[93] which these authorities can further specify and interpret as follows.

First, the GPC specification refers to the *selling or sharing* of personal information, which we interpret as types of *processing* under the GDPR.[94] While neither the selling nor sharing is explicitly mentioned in the GDPR's definition of processing, we consider them to constitute forms of *making available*, and thus to fall under the scope of GDPR's definition of processing.[95] The GDPR regulates broadly *any* processing of personal data with the enumerated forms of processing, "such as" collection, recording, and others serving as examples.[96] The use of selling or sharing within the GPC specification therefore falls squarely within the GDPR's definition of processing. On the other hand, forms of processing defined by the GDPR beyond making available, such as collection of personal data from the data subject, are not covered by the GPC specification.

Second, the GPC specification refers to *cross-context ad targeting*. While the specification does not define *context* or *cross-context*, it refers to the definition of context in the W3C Privacy Principles, according to which "[a] context is a physical or digital environment in which people interact with other actors, and which the people understand as distinct from other contexts."[97] Thus, the interpretation of a context depends on the expectations of data subjects and can refer to the first party websites or services that they are currently using. In such cases, *cross-context* means across parties or websites, even if they are owned by the same organisation or part of the same service, and cross-context behavioural targeting means using the data subject's activities across context to target ads.[98]

[90] Article 4(8) GDPR.
[91] The GPC specification mentions *processors* in §5.2, an informative section, as "instances where third party sharing may be permitted such as sharing to service providers/processors."
[92] We note that third parties under the GDPR may also qualify as joint controllers based on the factual influence of the parties, processing phases, and determination of purposes and means of processing, as evidenced in the jurisprudence of the CJEU, e.g., Case C-210/16 (*Wirtschaftsakademie Schleswig-Holstein*, 2018), Case C-40/17 (*FashionID*, 2019) and Case C-604/22 (*IAB TCF*, 2024).
[93] S. Zimmeck, P. Snyder, J. Brookman and A. Zucker-Scharff, 'Global Privacy Control (GPC) W3C Editor's Draft 3 April 2026' https://w3c.github.io/gpc/ accessed 21 April 2026.
[94] Article 4(2) GDPR regarding "processing of personal data" and Article 4(10) GDPR regarding third parties.
[95] Article 4(2) GDPR.
[96] Ibid.
[97] §A.2 in R. Berjon and J. Yasskin, 'Privacy Principles W3C Statement 15 May 2025' https://www.w3.org/TR/privacy-principles/#context accessed 21 April 2026.

The GPC specification intends to restrict data selling to or sharing with third parties, as well as cross-context ad targeting by a controller. However, GPC does not intend to limit the controller's other processing activities within the context of a data subject’s website visit. For example, if a controller conducts “tracking” and profiling of a data subject on its own site and *within the same context*, then this activity would not be restricted by GPC. In contrast, GPC would prevent third-party processing activities to the extent that these activities constitute data selling or sharing or cross-context ad targeting. We therefore consider the scope of GPC under the GDPR to be *inapplicable* to the processing conducted by first parties, i.e., controllers, when it is within the same context as the data subject’s website visit. We consider GPC to be *indiscriminate* regarding third parties as it prevents any and all processing considered *selling* or *sharing*.

We note that though the term context has no defined meaning for the GDPR, case law, such as *FashionID*,[99] can provide an interpretation of its determination for the GDPR, where joint-controller relationships can qualify as cross-context interactions. In *FashionID*, an online clothing retailer and a third party became joint-controllers for the initial act of data collection, after which they resumed their roles as controller and third party respectively. Additionally, their individual activities require a separation of control—such as through the data subject's consent—for a valid legal basis under the GDPR. This separation, though undefined in the GDPR, implies the existence and separation of contexts.

As GPC can only stop future data processing, it does not impact data processing that occurred before a data subject turned on GPC. However, any ongoing processing would still require a valid legal basis under the GDPR.

The GPC specification includes informative or *non-normative*, i.e., non-binding, guidance on how GPC may be interpreted under EU law,[100] referring to the potential application of rights, in particular, to the refusal and withdrawal of consent and the objection to legitimate interests under the GDPR. The drafters of GPC envision that if data subjects send a GPC signal, their request to prevent or stop processing of their personal data should be implemented through the relevant GDPR rights available for the used legal basis. This application follows from the GDPR requiring controllers to determine the appropriate legal basis for each data processing scenario, which in turn determines the choices available to data subjects to permit or restrict further processing.

---

[98] See also M. Veale and F.J. Zuiderveen Borgesius, ‘Adtech and Real-Time Bidding under European Data Protection Law’ (2022) 23(2) German Law Journal, pp. 226-256.
[99] See CJEU Case C-40/17 (*FashionID*, 2019) para.27 regarding user awareness and para.102 regarding ability to obtain consent, https://eur-lex.europa.eu/legal-content/EN/TXT/?uri=celex:62017CJ0040 accessed 21 April 2026.
[100] §5.2 GPC specification and §4.2 GPC Legal and Implementation Considerations Guide.

In the following subsections, we examine the use of GPC through illustrative scenarios reflecting common use cases based on actor's roles under the GDPR and experiences of data subjects. For each scenario, we discuss the most likely legal basis for the data processing and the resulting effects of using GPC under those bases. Our analysis mainly focuses on the legal bases of consent, contract performance, and legitimate interests.[101]

## 4.2 Scenario A: Third-party Ad Tracking and Consent as Legal Basis

In scenario A we consider the data subject visiting a website that integrates third-party advertising where third parties serve ads based on a profile they are building by tracking the data subject's activity across multiple websites, i.e., by performing behavioural advertising. Under the GDPR, the only legal basis for such behavioural advertising is the data subject's consent.[102] The GDPR requires that such consent is obtained *prior* to any processing, which means the website cannot perform behavioural advertising before the data subject has given consent. Further, as tracking and profiling for behavioural advertising typically relies on placing cookies on the data subject's device, it also requires consent under the ePD.[103] Thus, the current standard practice is that websites present a banner to data subjects requesting consent for behavioural advertising under both the GDPR and the ePD.

When the data subject visits the website and sends a GPC signal, the signal is communicated as part of every request made to access the website, including the first visit (see Figure 1). As the use of tracking and profiling by third parties for behavioural advertising qualifies as sharing data with a third party and as cross-context ad targeting, the GPC signal communicates the data subject's preference to prevent such activity. As the website receives the GPC signal before displaying any content, including a consent banner, it can take the GPC signal into account and adjust its behaviour regarding the consent request and the accompanying banner.[104] However, activities outside of the scope of GPC, such as data processing performed by the website itself for purposes of contextual advertising, remain unaffected.

If data subjects visit the same site a second time with an active GPC signal, their wish to prevent specific data processing for behavioural advertising remains the same as during the first visit. Thus, the use of GPC at each visit indicates a continued affirmation that the data subjects' wishes have not changed. However, if data subjects visit the site without an active GPC signal, such visits are a deviation from the initial visits where the GPC signal was communicated to the website. In this regard, the GPC specification does not require a website to keep a record of whether a data subject had enabled GPC in the past, i.e., it can simply check each request when it comes in for whether it contains a GPC signal. However, if a website does keep a record of the use of GPC, such as through a cookie,[105] it would be able to infer that a data subject has turned off GPC, which raises the question what turning off of GPC means. The GPC specification, intentionally, as mentioned in section 3, does not provide for any mechanism to communicate that a GPC signal is off and effectively requires browsers to not distinguish between not using GPC and disabling it.[106]

---

[101] Article 6(1)(a), 6(1)(b), and 6(1)(f) GDPR.

[102] Article 29 Data Protection Working Party, 'Opinion 06/2014 on the Notion of Legitimate Interests of the Data Controller under Article 7 of Directive 95/46/EC' (2014), pp. 46-47, https://ec.europa.eu/justice/article-29/documentation/opinion-recommendation/files/2014/wp217_en.pdf accessed 21 April 2026.

[103] Article 5(3) ePD.

[104] Current EU law does not prohibit the website from showing the data subject a consent banner, even if the website sees the data subject's GPC signal.

Under current EU law, the lack of a GPC signal, even if it was previously enabled, cannot be interpreted as giving consent, even to any previous processing to which the user refused or withdrew their consent. This means GPC signals can only convey the data subject's preference to prevent certain types of processing, and a lack of GPC signal or the act of disabling it does not imply *anything* in itself. Under Article 4(11) GDPR, consent requires an active unambiguous indication of a data subject's wishes, which is absent when no GPC signal is present or when it is deactivated. Therefore, if the data subject turns off their GPC signal after prior visits, the website cannot assume that the data subject has now given consent. While the website may view turning off GPC as an opportunity to request consent, both the GPC specification and the GDPR do not specify how or when controllers can ask for consent after the data subject has refused or withdrawn their consent, whether manually or through GPC.[107]

If a data subject consents to tracking-based advertising through a banner while the GPC signal is active, the validity of consent is unclear due to being ambiguous as per the GDPR. If we consider GPC as an indication of the data subject's wishes, then the consent that the data subject gives by clicking yes on the banner could be considered invalid due to the conflict between the automated GPC signal and the data subject's click. The website may argue that the data subject's choice in the banner represents an explicit, intentional choice by the user, whereas the GPC signal is automated and does not constitute a deliberate action. However, the GDPR treats ambiguity in consent strictly and does not establish a hierarchy to resolve conflicts between different expressions of choice.

If a data subject consented in a prior interaction without sending a GPC signal and is revisiting a site but now with an *active* GPC signal, the site now has an indication of the data subject's preferences through the use of GPC that the data subject no longer wishes to allow sharing data with third parties or allow cross-context ad targeting. This use of GPC *can* indicate the exercise of the data subject's right to withdraw consent. Under current law, the site may still decide to ask the data subject to confirm if they wish to withdraw their consent by showing a banner. On the other hand, the site may also consider that the data subject has already indicated the withdrawal by sending a GPC signal and not show a banner.

---

[105] Such use of cookies can be deemed necessary for the functioning of GPC and thus be exempt from consent under Article 5(3) ePD.

[106] From §3.2 of the GPC specification: "... and false if the person's preference was disabled or had not been set."

[107] While GDPR guidance suggests waiting a *reasonable* amount of time before requesting consent again, there is no fixed duration or period for doing so. Additionally, it is unclear whether the duration would reset with every GPC signal sent or would be counted from the first instance of the signal.

Without a legally binding interpretation or obligation, benevolent interpretations of GPC that reduce banners thus rely on the goodwill of websites and controllers. For the benefits from GPC to materialise, relevant regulations would need to clarify when websites may or may not use banners in an explicit and enforceable manner. For example, the law could state that if a site is aware that the data subject has indicated the withdrawal by sending a GPC signal, the site is not permitted to show a banner for the activities covered by the consent withdrawal.[108] Similarly, for automated signals to be useful in general, regulations would also have to clarify potential conflicts between the GPC signal and choices made in banners, or between multiple different signals being used simultaneously.[109]

Scenario A illustrates that when consent serves as the legal basis for data processing, GPC could alleviate data subjects from refusing or withdrawing consent via banners. But without legal intervention in the EU, websites can continue to show consent banners to data subjects. Moreover, as GPC is limited in scope to third party data sharing and a controller's cross-context ad targeting, it can not reduce banners for other purposes, such as requests to consent to functional cookies.

## 4.3 Scenario B: Processor Payment Service and Necessity for Contract Performance as Legal Basis

In scenario B we consider a website that uses a processor to integrate payment services, allowing the data subject to make payments for their use of the website. As explained in section 4.1, we interpret the GPC specification such that entities acting as processors under the GDPR are not third parties for the purposes of GPC. Consequently, because no third parties are involved in this processing, and assuming this does not involve any cross-context ad targeting by the controller, the use of GPC has no effect in this scenario.

Alternatively, even if we assume that processors are treated as third parties under the GPC specification, GPC would still have no effect because the data sharing with the payment service would be covered by the legal basis of necessity for contract performance. The processing of personal data would be necessary to fulfil a contract with the data subject.[110] This scenario illustrates the broader point that GPC is generally not applicable if contract performance serves as the legal basis.

## 4.4 Scenario C: Third Parties and Legitimate Interests as Legal Basis

In scenario C, as in scenario B, a website shares data with a payment service provider. However, in this scenario the payment service provider additionally conducts its own processing to prevent and detect scams and fraud. As the payment service provider solely determines the purpose and means of processing for these activities, it qualifies as a data controller under the GDPR for those processing activities.[111] Since the website, as the initial controller, has authorised the sharing of personal data, the payment service provider is also a third party for the processing activities of the website.

---

[108] Recital (26) ePD (“withdraw his/her consent to such processing”) and Article 7(3) GDPR (“The data subject shall have the right to withdraw his or her consent at any time.”).

[109] For evidence of ambiguity between banner choices and privacy signals, see M. Hils, D.W. Woods and R. Böhme, ‘Conflicting Privacy Preference Signals in the Wild’ (2021) https://arxiv.org/abs/2109.14286 accessed 21 April 2026.

[110] Article 6(1)(b) GDPR.

In this scenario we assume that the payment service provider relies on legitimate interests as the legal basis for its own processing activities.[112] Thus, the payment service provider is obligated under the GDPR to conduct a three-tiered balancing test to ensure that (1) the interests pursued are legitimate, (2) the processing of personal data is necessary, and (3) the processing does not override the interests or fundamental rights and freedoms of the data subject.[113] We assume that the payment service provider satisfies the requirements of the balancing test. We also assume that it complies with other relevant obligations, for instance, regarding transparency.[114]

If the data subject visits the website with an active GPC signal, the site can interpret it as an indication that the data subject wants to stop it from sharing data with third parties—namely, the payment service provider.[115] Since the legal basis used for sharing data is the legitimate interests provision, the GDPR grants data subjects the right to object,[116] including via "automated means using technical specifications."[117] GPC principally meets this criterion and, thus, can serve as an indication that the data subject is exercising the right to object. As a practical matter, since the text of the GDPR does not specify which automated means controllers are required to accept, the use of GPC in this regard depends on EU legislators' and regulators' guidance to mandate websites to recognise and act on the GPC signals they receive.

Assuming that GPC is a valid means to exercise the right to object, the GDPR allows controllers and third parties to refuse the objection by demonstrating compelling legitimate grounds that override the data subject's interests, rights, and freedoms.[118] The standard for whether to refuse an objection is high. If a data subject has invoked the right to object, it is not sufficient for the controller or the third party to just demonstrate that its earlier legitimate interests assessment regarding that processing was correct.[119] For legitimate grounds to be *compelling* they should be essential to the controller or to the third party in whose legitimate interests the data are being processed.[120] Examples of such cases are controllers being compelled to process the personal data in order to protect its organisation or systems from serious immediate harm or from a severe penalty which would seriously affect its business.[121] In contrast, showing that the processing would simply be beneficial or advantageous to the controller would not necessarily meet this threshold.[122]

---

[111] European Data Protection Board, 'Guidelines 07/2020 on the concepts of controller and processor in the GDPR' Version 2.1 (2022) https://www.edpb.europa.eu/system/files_en?file=2023-10/EDPB_guidelines_202007_controllerprocessor_final_en.pdf accessed 21 April 2026.

[112] Article 6(1)(f) GDPR; Recital 47 GDPR.

[113] Article 6(1)(f) GDPR; European Data Protection Board, 'Guidelines 1/2024 on processing of personal data based on Article 6(1)(f) GDPR' Version 1.0 (2024) https://www.edpb.europa.eu/system/files/2024-10/edpb_guidelines_202401_legitimateinterest_en.pdf accessed 21 April 2026.

[114] Article 14 GDPR regarding provision of notice when data not collected from the data subject, and European Data Protection Board, 'Guidelines 07/2020 on the concepts of controller and processor in the GDPR' Version 2.1 (2022) https://www.edpb.europa.eu/system/files_en?file=2023-10/EDPB_guidelines_202007_controllerprocessor_final_en.pdf accessed 21 April 2026.

[115] Selling and cross-contextual ad targeting are not relevant for preventing and detecting scams and fraud.

[116] Article 21(1) GDPR.

[117] Article 21(5) GDPR.

The effects of using GPC to prevent processing thus take effect only in cases where a controller as a third party lacks compelling legitimate grounds to override a data subject's right to object. However, in our scenario, the payment processor could argue that processing personal data is essential to detect fraud and prevent severe financial losses. If these grounds are deemed compelling—which they may well be—they would override the data subject's objection. Consequently, in such scenarios, sending GPC signals as a data subject's objection could result in the objection being refused by the controller or third party, and a requirement to demonstrate the grounds for the refusal to the data subject.[123] As the GPC specification does not provide for a mechanism to indicate acceptance or refusal of a right, the payment service provider may resort to banners, which would limit the effectiveness of GPC in reducing banners.

The payment service provider acting as the third party is also a processor for the website illustrating that one entity can take on multiple roles under the GDPR for different processing purposes. In such cases, the use of GPC may not result in stopping the sharing, as the processor (the payment service provider) may require the data to complete its contractual obligations with the controller (the website). Thus, given that the scope of the GPC signal is limited to preventing data sharing with third parties, GPC may not always have an impact on the further processing of data shared with a third party that also acts as a processor. However, to the extent processing activities can be cleanly separated, GPC can have different effects on the separated processing activities.

---

[118] Article 21(1) GDPR, and §4 European Data Protection Board, 'Guidelines 1/2024 on processing of personal data based on Article 6(1)(f)' GDPR Version 1.0 (2024) https://www.edpb.europa.eu/system/files/2024-10/edpb_guidelines_202401_legitimateinterest_en.pdf accessed 21 April 2026.

[119] Article 21(1) GDPR, and §4 European Data Protection Board, 'Guidelines 1/2024 on processing of personal data based on Article 6(1)(f) GDPR' Version 1.0, para. 73 (2024) https://www.edpb.europa.eu/system/files/2024-10/edpb_guidelines_202401_legitimateinterest_en.pdf accessed 21 April 2026.

[120] Ibid.

[121] Ibid.

[122] Ibid.

[123] While Article 21(1) GDPR does not specify to whom the justification must be provided, we interpret from the EDPB's guidelines that it is the data subject. See European Data Protection Board, 'Guidelines 1/2024 on processing of personal data based on Article 6(1)(f) GDPR' Version 1.0, para. 73 (2024) https://www.edpb.europa.eu/system/files/2024-10/edpb_guidelines_202401_legitimateinterest_en.pdf accessed 21 April 2026 ("If the controller fails to provide such proof, the data subject is entitled to request the erasure of the data …").

In addition to being a third party and processor, the payment service provider would also likely be a joint controller under the GDPR.[124] As discussed, the GPC specification is focused on a single entity acting as the first party.[125] In this regard, the GPC specification implies that the user's intent governs the determination of who should be the first party when multiple parties are involved. However, the GDPR does not differentiate between the controllers involved in a joint controller relationship as first and third parties but instead requires them to determine which responsibilities should be shared.[126] The GDPR's notion of responsibilities also goes beyond user-facing activities based on the assumption that entities exerting substantial influence on determining the means and purposes without participating in the processing can also qualify as joint controllers.[127] If the activities of both joint controllers in this scenario are within the context the data subject expects—which they may well be—GPC would not apply. However, if the activities of the two joint controllers do not fall within the same context, a determination of first and third parties becomes necessary for GPC to prevent processing.[128]

Scenario C illustrates that GPC can be a mechanism for data subjects to exercise their right to object under the GDPR where a third party relies on the legitimate interests provision as a legal basis. However, it also shows the need for a nuanced evaluation to identify which practices are covered by GPC and which are out of scope. Such an evaluation is especially important as data subjects may otherwise become inundated with banners, for example, explaining why their objection was refused. It also shows that applying GPC to the GDPR in joint controller scenarios requires a case by case analysis to determine which joint controller activities would be subject to GPC's scope and application.

## 4.5 Scenario D: Services Using Cookies and User Device Storage

In scenario D we take a closer look at the relevance of GPC for the ePD.[129] Article 5(3) ePD concerns the *storage* or *access* of *information*, for example, in a cookie, on the user's device.[130] Article 5(3) ePD requires any party that stores or reads a cookie on a user's device to obtain that user's prior consent, after giving that user clear and comprehensive information in accordance with the GDPR, inter alia, about the purposes of the processing. The requirements for valid consent are strict and are defined in the GDPR. The party does not have to obtain the user's consent if it uses the cookie to carry out the transmission of a communication (for instance, a cookie for a log-in procedure) or is strictly necessary to provide a service explicitly

[124] Ibid. 87.
[125] Section 3.2.
[126] Section 4.1.
[127] Ibid. 87.
[128] As discussed previously (specifically in section 4.1), this determination is a complex procedure due to the GDPR's broad interpretation of controllers as well as joint-controllers. It requires a further in-depth analysis of joint-controllers based on GPC's notion of *context* and *user intent* and how they would need to be determined on a case-by-case basis, which, however, is outside of the scope of our study.
[129] To the extent that personal data is processed, the GDPR would apply as well.
[130] In scenario D, we use *cookies* and *information* synonymously.

requested by the user (for instance, a cookie that is used to implement a digital shopping cart).[131]

To understand the applicability of GPC to the eDP we focus on the GPC signal indicating a refusal to the storage and access on the user's device. As the GPC signal represents the user's preference to not allow certain types of processing, it can act as a refusal in the context of the ePD (and as a refusal to consent under the GDPR). However, per the GPC specification such refusal only applies to the selling or sharing of personal data with third parties and cross-context ad targeting by any party. Thus, the use of GPC can only restrict some operations under the ePD.

To illustrate its application, consider a website storing and using a cookie on the user's device containing an identifier for the user. The GPC signal will prohibit the website from selling or sharing information that it gathers through the cookie with a third party, or for cross-context ad targeting by any party. GPC will not have any effect for other uses of the cookies—such as tracking the user across the website (assuming it falls under the same *context*). For such cookies that are not in scope of GPC, consent is still needed under the ePD. This means if the website places a cookie on the user's computer it must ask the user for consent, even if the site does not share data (that it gathers through the cookie) with third parties. Similarly, the ePD still requires consent if a website enables a third party to place a cookie, even if the third party chooses not to collect data through the cookie.

Thus, at least for purposes not in scope of GPC, the consent requirements of the ePD still necessitate a banner. However, as discussed previously (specifically section 4.2), even if banners whose purposes are all within the scope of GPC would not need to be shown, additional legislative measures may be required to ensure this outcome. The use of GPC thus has limited benefits under the ePD, especially regarding prevention of cookie banners.

## 4.6 Irrelevance of Non-Commercial Legal Bases

We focus our evaluation on the legal bases that are most relevant in the online ecosystem from a commercial perspective—consent, contract, and legitimate interests—because GPC is technically and functionally designed to address the voluntary selling or sharing of data and cross-context ad targeting. Consequently, the remaining legal bases—legal obligation,[132] vital interests,[133] and public interest or official authority[134]—are less relevant here.

Processing grounded in a legal obligation is mandated by statute; an automated objection signal cannot legally nullify a controller's duty to comply with the law, e.g., retaining transaction records for tax compliance or fraud investigations. Similarly, processing necessary to protect vital interests typically involves life-critical emergencies where automated commercial privacy signals are neither practical nor legally pertinent. While data subjects have a right to object to processing based on public interest or official authority under Article 21(1) GDPR, GPC is irrelevant as controllers are not carrying out tasks in the public interest or in the exercise of official authority.

[131] Art. 5(3) ePD. See also European Data Protection Board, 'Guidelines 2/2023 on Technical Scope of Art. 5(3) of ePrivacy Directive' Version 2.0 (2024) https://www.edpb.europa.eu/system/files/2024-10/edpb_guidelines_202302_technical_scope_art_53_eprivacydirective_v2_en_0.pdf accessed 21 April 2026.
[132] Article 6(1)(c) GDPR.
[133] Article 6(1)(d) GDPR.
[134] Article 6(1)(e) GDPR.

## 4.7 GPC under the Proposed Article 88b GDPR

The proposed Article 88b(1) GDPR[135] defines a technical signal capable of giving and rejecting consent and of objecting to processing based on the legitimate interests provision. While Article 88b GDPR does not specify what is required for an automated expression of consent to be valid, we consider that it must, at least, meet the existing GDPR consent requirements. In addition, we also consider that the technical signal must support withdrawal following the GDPR’s requirement for withdrawing consent to be, at least, as easy as giving it.[136] As discussed in section 4.2, the use of a GPC signal cannot indicate a data subject’s unambiguous consent. However, a GPC signal can indicate consent rejection and withdrawal and an objection to legitimate interests. The scope of the GPC signal would apply to processing activities that constitute selling or sharing with third parties or a controller's cross-context ad targeting. This interpretation results in an application of GPC that does not cover the full scope of the GDPR's requirements, in particular regarding giving consent and the granularity of purposes,[137] and, by extension, the requirements of Article 88b GDPR.

The interplay between GPC and EU law raises more questions that go beyond the scope of our evaluation. We mention some suggestions for further research. For example, can GPC be an acceptable implementation under Article 88b(4) GDPR, even though GPC is developed at the W3C, which is not a recognised European standardisation body?[138] Another question arises from the fact that GPC does not cover the full scope of requirements listed in Article 88b(1) GDPR, since GPC is only applicable to a few processing purposes. By referring to “standards” in the plural, the provision seems to assume that there can be more than one standard, which suggests a possibility to have different complimentary implementations. Similarly, another relevant question is how a GPC signal would work with Article 88a GDPR, which is also proposed in the Digital Omnibus and introduces limitations on consent requirements for certain cookies.[139]

---

[135] Ibid. 1.
[136] Article 7(3) GDPR.
[137] See European Data Protection Board, ‘Guidelines 05/2020 on consent under Regulation 2016/679’ Version 1.1, para. 89 (2020) https://www.edpb.europa.eu/sites/default/files/files/file1/edpb_guidelines_202005_consent_en.pdf accessed 21 April 2026: “Such settings should be developed in line with the conditions for valid consent in the GDPR, as for instance that the consent shall be granular for each of the envisaged purposes … .”
[138] While there is a path for ratification of external standards with minimal modifications through a recognised European standardisation body, such as commonly done for ISO and IEC standards, our question is based on the nature of GPC's development occurring outside recognised bodies.
[139] The exemptions in Article 88a(1) GDPR are limited to only a few purposes and only cover cookies, which leaves many common purposes, such as profiling-based measurements or third-party tracking, and

It is also unclear how Article 88b GDPR envisions existing mechanisms that may already be present in the data subject's device or browser should be interpreted, such as GPC. If they fall under the provision, what should be the path to ratification? If GPC does not fall under the provision, how would this omission affect other device or browser settings and configurations, for example, regarding blocking of third party cookies? We also raise the question of cases where more than one standard with partially overlapping scope, such as GPC and a hypothetical Article 88b GDPR implementation, are applied at the same time and how the resolution of conflicts between them is addressed.

Per Article 88b(3) GDPR, GPC would not apply to controllers that are *media service providers* when providing a media service. Presumably the provision seeks to safeguard the economic viability of the media. That is a noble goal, but the exception leaves open questions and brings risks. The provision does not clarify if it also exempts third-party behavioural advertising used by media providers or would only permit first-party advertising implemented by the media provider itself. From a privacy and data protection perspective, by exempting media service providers from automated indications of data subjects' choices, the provision risks undermining the privacy of, for instance, readers of online newspapers.

# 5 Discussion and Future Directions

We evaluated GPC as a technical mechanism standardised at the W3C for its application under the current EU data protection framework as well as the proposed Article 88b GDPR under the Digital Omnibus. Our evaluation shows that while GPC can be aligned in some respects with the EU framework, open questions remain. The GPC specification's definitions are strongly influenced by US state privacy laws and do not map neatly to the EU framework. Consequently, without updates to the specification and its guide and to EU legislation, GPC would not be able to achieve its promise of automated rights exercise in the EU.

GPC should be viewed not as a rigid standard, but as an "empty canvas" with infrastructure in browsers and on websites already in place that is ready for legislators and regulators to use to improve data protection in the EU. Equally, the ongoing development of GPC presents an opportunity for the standard editors to evolve GPC towards application in the EU. Alongside developing homegrown and idealised standards in the future, as envisioned in the proposed Article 88b GDPR, EU legislators and regulators also have the authority to map the existing, jurisdiction-agnostic GPC signal to a specific set of practices and rights under EU law. This approach is pragmatic considering that GPC is already adopted by multiple stakeholders—most importantly browsers vendors, consent management platforms, and media publishers, is enforceable in several US jurisdictions, and is required as a mandatory browser setting per California law from 1 January 2027. The GPC specification does not govern the law; rather, the law governs how the signal must be honoured. Therefore, EU authorities can define what GPC

---

use of technologies other than cookies, outside the scope of its effects. In this paper, we do not further discuss Article 88a GDPR due to length constraints, and focus our analysis on Article 88b GDPR.

as a “Do Not Process” signal means in the context of EU data protection law. It is their opportunity to fill the empty canvas with the legal requirements in the EU.

As it stands, GPC’s reduction of consent banners relies on the goodwill of website operators and other controllers. Under current law, after all, websites could still show consent banners to people who express with GPC that they do not want the website to share their data with others. Therefore, for GPC’s benefits to materialise, relevant laws, regulations, or official legal guidance would need to mandate its use and specify its effects in an explicit and enforceable manner.

If adopted, the Digital Omnibus could give automated signals like GPC legal enforceability under the GDPR. To support the development, our recommendations for the use of GPC in the EU are as follows:

- The proposed Article 88b GDPR should provide for existing standards, such as GPC, to be enforceable for restricting processing of personal data in some situations.
- GDPR authorities should clarify the applicability and enforceability of GPC by providing guidance as follows: (1) Where the legal basis is consent, data subjects who broadcast GPC are refusing or withdrawing consent under Article 7(3) GDPR to the use of their personal data for cross-context ad targeting, i.e., websites (controllers) are prohibited to share data subjects’ personal data for this purpose with third parties. (2) Where the legal basis is legitimate interest, data subjects who broadcast GPC signals are objecting under Article 21(5) GDPR to their personal data being used for cross-context ad targeting.
- To reduce consent fatigue, EU legislators or regulators should provide guidance as to when controllers can and cannot display consent banners when data subjects exercise their rights through existing standards, such as GPC.

The experiences of GPC adoption and enforcement in the US offer valuable insights for the EU, namely that technical signals require legal backing to be effective. These experiences also include exemplars such as the legal obligation for browsers to, effectively, implement GPC under the Opt Me Out Act per California Assembly Bill 566 starting from 1 January 2027[140] and a demonstrated positive outcome such as a reduction in tracking cookies.[141] Successful implementation of EU rights mechanisms thus requires a closer integration of law and technology. Integrating automated signals like GPC into the EU’s regulatory framework could bridge the gap between the GDPR’s robust substantive protections and the procedural ease of automating privacy rights. Such a step would not only alleviate the burden of manual consent management for EU residents but also foster greater international harmonization in the technical exercise of privacy rights.

---

[140] California State Legislature, AB-566 California Consumer Privacy Act of 2018: opt-out preference signal, https://leginfo.legislature.ca.gov/faces/billNavClient.xhtml?bill_id=202520260AB566.
[141] A. Rasaii et al., ‘Intractable Cookie Crumbs: Unveiling the Nexus of Stateful Banner Interaction and Tracking Cookies’ (2025) Proc. Privacy Enhancing Technologies (PoPETs) https://doi.org/10.56553/popets-2025-0138 accessed 21 April 2026.

In this study, we focused on the interplay between GPC and the GDPR and ePD. We give some suggestions for future research. For example, the use of GPC impacts *first* and *third parties*—and potentially larger and smaller companies—differently. As GPC leaves most first-party processing intact while preventing third-party processing, it could create an advantage for existing larger platforms and popular websites. There could also be an opposite effect as banner-level consent mechanisms can advantage incumbents, as users may be more likely to click accept to consent banners on popular websites.[142] GPC could thus also be analysed through an economic lens regarding the interplay between regulation such as competition law, the Digital Services Act (DSA), the Digital Markets Act, and consumer protection law.

We began our inquiry with the question whether the GPC standard can limit the number of consent banners in the EU. Our answer is: partly and possibly more in the future. If a website only implements tracking cookies by third parties and necessary cookies (for which no consent is needed), the site does not need to show a consent banner if a data subject signals, with GPC, their consent refusal or withdrawal or their objection to cross-context ad targeting. Under current law, the website could honour the GPC signal and refrain from placing the tracking cookies. A website could also choose to place fewer cookies for analytics when it sees a visitor's GPC signal, and thus to show fewer consent banners.

The application of GPC as suggested here is likely in line with the majority of users' wishes.[143] Thus, it can be a pragmatic starting point towards a more nuanced and comprehensive solution. While GPC may not eliminate consent banners—particularly, given the requirements of the ePD—we think it can reduce the burden on users over time as EU authorities develop and refine their interpretation of GPC and similar standards.

# 6 Conclusion

This paper explored to which extent GPC can be applied in the EU to enable data subjects to express their wishes around tracking and data collection online. In principle, automated signals along the lines of GPC are a good idea, because they can make expressing or refusing consent user-friendlier. Earlier attempts to develop such signals, for instance Do Not Track, largely failed because websites and other companies were not legally required to honour such signals. In some US states, there is now a legal requirement to honour a GPC signal. The legal backing of GPC is thus an important development.

[142] C. Farronato, A. Fradkin and T. Lin, 'Designing Consent: Choice Architecture and Consumer Welfare in Data Sharing' (2025) NBER Working Paper 34025, https://www.nber.org/papers/w34025 accessed 21 April 2026.

[143] Notably, a recent study showed that among users who were forced to make choices, 83% accepted functional/preference cookies, while only 7% consented to ad selection, delivery, and reporting. See C. Farronato, A. Fradkin and T. Lin, 'Designing Consent: Choice Architecture and Consumer Welfare in Data Sharing' (2025) NBER Working Paper 34025, https://www.nber.org/papers/w34025 accessed 21 April 2026. See also L. Kyi et al., '"It doesn't tell me anything about how my data is used": User Perceptions of Data Collection Purposes' (2024) ACM Proc. CHI Conference on Human Factors in Computing Systems (CHI), Article 984, pp. 1-12, https://doi.org/10.1145/3613904.3642260 accessed 21 April 2026. In this interview study most of the 23 participants preferred not to share data for personalized advertising and advertising. purposes. They were more comfortable sharing their data for strictly necessary, performance and functionality, and statistics and analytics purposes.

What could GPC mean for internet users in the EU? In the short term, well-meaning websites could already stop sharing data about their visitors with third parties when they receive a GPC signal, even though EU law does not require that website reaction at the moment. However, we also identified some challenges with applying GPC in the EU. First, in many situations, EU law (in particular the ePD) requires websites to ask for consent for the placing of cookies, even if a website honours GPC signals. Hence, in such cases GPC hardly reduces consent banners. Second, we showed that the wording used in the GPC standard (such as *first party* and *selling* or *sharing*) does not neatly map to legal terms in the GDPR. In the short term, EU regulators could clarify how websites should interpret GPC signals. We described a reasonable interpretation, broadly summarised: GPC should mean refusing and withdrawing consent, and objecting to processing if the processing involves sharing data with third parties or for advertising across a controller's various services. Third, we showed that GPC is primarily meant to limit a particular activity: cross-context ad targeting, and more broadly, the sharing or selling user information to third parties. Therefore, GPC does not say anything about many data collection practices that users may dislike, such as “tracking” or profiling by the website itself, which means that even if websites honour GPC signals as envisioned by the GPC specification, they can continue other practices not in scope of GPC.

In the long term, EU legislators and regulators could cooperate with the W3C or other standard bodies to combine user-friendly privacy signals with regulation to ease the process of lawmaking and enforcement related to the internet. GPC illustrates that the law and technical standards can supplement each other. The EU legislator can thus take inspiration from developments around GPC in its own efforts.

## Acknowledgments and Declaration of Competing Interest

We thank Nataliia Bielova for her comments on an earlier version of our paper. Sebastian Zimmeck has received funding from the National Science Foundation (Award #2055196), Wesleyan University, its Department of Mathematics and Computer Science, and the Anil Fernando Endowment. Harshvardhan J. Pandit is part of the AI Accountability Lab, which has been funded under the AI Collaborative, an Initiative of the Omidyar Group, the Bestseller Foundation, the AI Security Institute (AISI), and the John D. and Catherine T. MacArthur Foundation. It is part of the ADAPT Centre for Digital Media Technology that is funded by Research Ireland through the Research Centres Programme and is co-funded under the European Regional Development Fund (ERDF) through Grant 13/RC/2106_P2 and has received funding through the RECITALS project funded by the European Commission's Horizon research and innovation programme under grant agreement No.101168490. Cristiana Santos was supported by the Utrecht Centre for Regulation and Enforcement in Europe (RENFORCE). Konrad Kollnig has received funding from the Dutch National Growth Fund.

The authors declare that they have no known competing financial interests or personal relationships that could have appeared to influence the work reported in this paper. Sebastian Zimmeck is a current editor of the GPC specification. Robin Berjon is a former editor of the GPC specification. Konrad Kollnig has a consulting/advisory relationship with the Open Data Institute.